\documentclass[apj]{emulateapj}
\usepackage{apjfonts}
\usepackage{graphicx}
\usepackage{color}
\usepackage{amsbsy}
\usepackage{mathpazo}
\usepackage{multirow}

\newlength{\hfwidth}
\newlength{\hfwidthsingle}
\newcommand{\pderiv}[2]{\frac{\partial{#1}}{\partial{#2}}}

\newcommand{\ttimes}[1]{10^{#1}}
\newcommand{\xtimes}[2]{#1\times{10^{#2}}}

\renewcommand{\v}[1]{{\boldsymbol{#1}}} %for vectors

\newcommand{\advec}{\left(\v{u}\cdot\del\right)}

\newcommand{\del}{\v{\nabla}}

\newcommand{\Div}{\del\cdot}
\newcommand{\curl}{\del\times}

\newcommand{\va}{v_{_{\rm A}}}
\newcommand{\vao}{v_{_{\rm A0}}}

\newcommand{\Eq}[1]{Eq.~(\ref{#1})}

\newcommand{\eq}[1]{\Eq{#1}}

\newcommand{\Fig}[1]{Fig.~\ref{#1}}
\newcommand{\fig}[1]{\Fig{#1}}

\shorttitle{RWI in MHD}
\shortauthors{Lyra \& Mac Low}

\slugcomment{Draft version}

\begin{document}

\title{Rossby wave instability at dead zone boundaries \\
in 3D resistive magnetohydrodynamical global models of protoplanetary disks}

\author{Wladimir Lyra\altaffilmark{1,2,3} and Mordecai-Mark Mac Low\altaffilmark{1}}
\email{wlyra@jpl.nasa.gov,\ mordecai@amnh.org}
\altaffiltext{1}{Department of Astrophysics, American Museum of Natural History, 79th Street at Central Park West, New York, NY, 10024, USA}
\altaffiltext{2}{Jet Propulsion Laboratory, California Institute of Technology, 4800 Oak Grove Drive, Pasadena, CA, 91109, USA}
\altaffiltext{3}{NASA Carl Sagan Fellow}

%\date{Received ; Accepted}

\begin{abstract}
It has been suggested that the transition between magnetorotationally active
and dead zones in protoplanetary disks should be prone to the excitation of
vortices via Rossby wave instability (RWI). However, the only
numerical evidence for this has come from alpha disk models,
where the magnetic field evolution is not followed, and the effect of
turbulence is parametrized by Laplacian viscosity. We aim to establish
the phenomenology of the flow in the transition in 3D
resistive-magnetohydrodynamical models. We model the transition by a
sharp jump in resistivity, as expected in the inner dead zone boundary,
using the {\sc Pencil Code} to simulate the flow. We find that vortices are
readily excited in the dead side of the transition. We measure the mass
accretion rate finding similar levels of Reynolds stress at the dead and
active zones, at the $\alpha\approx\ttimes{-2}$ level.
The vortex sits in a pressure maximum and
does not migrate, surviving until the end of the simulation. A pressure
maximum in the active zone also triggers the RWI. The magnetized vortex
that results should be disrupted by parasitical magneto-elliptic instabilities, 
yet it subsists in high resolution. This suggests that either the parasitic modes are still
numerically damped, or that the RWI supplies vorticity faster than they
can destroy it. We conclude that the resistive transition between
the active and dead zones in the inner regions of protoplanetary disks,
if sharp enough, can indeed excite vortices via RWI. Our results lend
credence to previous works that relied on the alpha-disk approximation, and
caution against the use of overly reduced azimuthal coverage on modeling
this transition.
\end{abstract}

\section{Introduction}
\label{sect:introduction}

The formation of planets remains one of the most challenging problems 
of contemporary astrophysics. The current paradigm in planet formation 
theory describes a hierarchical growth of solid bodies, 
from interstellar dust grains to rocky planetary cores 
\citep{Safronov,Lyttleton,Goldreich&Ward,Youdin&Shu}. A 
particularly difficult phase in the process is the growth from centimeter 
sized pebbles and meter-sized boulders to planetary embryos the size of our Moon or 
Mars. Objects in the pebble to boulder range are expected to drift 
inward extremely rapidly in a protoplanetary disk, so that they would 
generally fall into the central star well before larger bodies can form 
by simple accumulation \citep{Weidenschilling,Brauer}. 

Ways to bypass this problem have focused on inhomogeneities in the flow, in order 
to trap particles in their migrating path. \citet{Cuzzi} proposed a model in 
which mm-sized particles are trapped in the smallest eddies in the flow, forming 
``sandpile'' planetesimals in the 10-100 km range (though that model has been 
criticized by \citealt{Chang&Oishi} and \citealt{Pan}). Particles may also be trapped in 
mesoscale ``zonal flows'' \citep{Lyra08a,Johansen09,Simon12} 
that are local inversions in the angular velocity profile, brought about by spatial variations in 
magnetic pressure. The particles themselves can give rise to the necessary inhomogeneities, 
as their migrating streaming flow develops into a traffic-jam instability 
\citep{Youdin&Shu,Youdin&Goodman,Youdin&Johansen}, 
leading to intense particle clumping \citep{Johansen&Youdin} 
and subsequent planetesimal formation at the 
Ceres-mass range \citep{Johansen07}. It has also been recently 
proposed that icy planetesimals may form from direct coagulation 
\citep{Okuzumi} due to the enhanced sticking properties of ices.

Another process has been suggested, that combines several of the advantages (as 
well as many of the problems) of the scenarios described above, and is the 
subject of this work. Turbulence in the largest scales of the flow, 
in the form of large scale vortices, has been 
independently proposed by \citet{Barge&Sommeria} and 
\citet{Tanga} as fast routes for planet formation, for two main reasons. 
First, vortices are equilibrium solutions of the Navier-Stokes equations, and 
thus are persistent structures 
in hydrodynamic flows, as seen in the Great Red Spot of Jupiter, a 
remarkable high pressure vortex stable since first spotted, over 
three hundred years ago \citep{Hooke,Cassini}{\footnote{The reader is
refereed to the fascinating account of \citet{Falorni} on 
the history of the discovery of the Spot, where the author debates 
the claims of primacy to Hooke or Cassini.}}. The second is that the 
equilibrium is geostrophic, i.e, between 
the Coriolis force and the pressure gradient force. As solids do not 
feel the pressure force, the Coriolis force will lead them out of the 
vortex if cyclonic and into the eye if anticyclonic. As the shear 
enforces that only anticyclonic vortices persist 
\citep{Marcus,Adams&Watkins,Bracco,Godon&Livio99}, this becomes a very 
effective mechanism to concentrate solid particles \citep{Klahr06}, as also 
observed in numerical simulations \citep{Godon&Livio00,Johansen04,Fromang&Nelson05,Inaba&Barge}. 
It was further shown by \citet{Lyra08b,Lyra09} that in the limit 
where turbulence in the vortex core is absent and the fragmentation of 
particles is ignored, the concentration of solids easily reaches the 
conditions necessary to gravitationally collapse them into planets. 

Nevertheless, as exciting as the vortex hypothesis may be, for a long time no 
plausible mechanism for the formation and sustenance of such storm systems in 
disks could be found. It is well known that three-dimensional vortices fall 
prey to elliptical instabilities \citep{Bayly,Pierrehumbert,Kerswell,Lesur&Papaloizou}, 
a general name given to parasitic instabilities of closed elliptical 
streamlines \citep[see e.g. the review of][]{Kerswell}.
Fluids in rigid rotation support a spectrum 
of stable inertial waves, the simplest ones being circularly polarized transverse 
plane waves oscillating at twice the base frequency \citep[see e.g.][]{Chandrasekhar61}. 
Strain is introduced when the motion passes from circular to elliptical, and 
some three-dimensional modes find resonance with the underlying strain field.  
Because of its intrinsic three-dimensional character, the absence of elliptic 
instability modifies the turbulent energy cascade from direct to inverse 
in 2D \citep{Batchelor}. The result is that the series of viscous mergings that turn 
small and mesoscale eddies into large vortices in the integral scale is not 
expected to occur in systems that significantly depart from two-dimensionality. 
In other words, vortices do not spontaneously develop in three dimensions from 
initial small scale noise as they do in two dimensions, and must therefore be the 
result of some instability. Because the Keplerian shear provides stabilization 
of linear axisymmetric disturbances at all Reynolds numbers \citep[Rayleigh criterion;][]{Strutt80,Strutt16}, in 
non-magnetized disks any such instability must be either non-axisymmetric, 
non-linear, or rely on a modification of the angular velocity profile to 
circumvent the stabilizing effect of the shear.

Based on these ideas, two main processes have been proposed for the formation 
of vortices in disks. One is the non-linear radial convection process that has 
been referred to as global baroclinic instability \citep{Klahr&Bodenheimer,Klahr04}, 
subcritical baroclinic instability 
\citep{Lesur&Papaloizou,Paardekooper}, and simply 
baroclinic instability \citep{Lyra&Klahr}. This shall not be 
dealt with in this barotropic paper. We will focus here on what 
has become known as Rossby wave instability \citep[RWI;][]{Lovelace,Li00,Li01,Umurhan}. 

The RWI relies on a modification of the angular velocity profile, brought 
about by a local extremum of an entropy-modified 
potential vorticity quantity (or simply the potential 
vorticity in the case of a barotropic flow). The 
extremum in potential vorticity launches inertial-acoustic waves, akin to 
Rossby waves in planetary atmospheres, and traps modes in its co-rotational 
singularity \citep[see fig 1 of][]{Meheut10}. The mechanism was first 
discussed by \citet{Lovelace&Hohlfeld} in the context of self-gravitating 
galactic disks, and mentioned en passant by \citet{Toomre}, who called them 
``edge modes''. The instability was discussed again by 
\citet{Papaloizou&Pringle84,Papaloizou&Pringle85} in the context of so-called 
``slender torus'' models, hot disks around quasars modeled locally in radius. 
The first non-linear numerical simulation of the instability, done by \citet{Hawley},
found that the trapped non-axisymmetric modes evolved into anticyclonic vortices, 
but could not follow the calculation until full saturation. 
The interest on the instability waned after the 
re-discovery of the magneto-rotational instability \citep[MRI;][]{Velikhov,Chandrasekhar60}
by \citet{Balbus&Hawley} and its rise to paradigmatic status as the source of angular 
momentum transport and turbulence in disks. The RWI was re-analyzed 
by \citet{Lovelace}, who expanded the linear analysis of \citet{Papaloizou&Pringle84} 
to non-barotropic disks, and gave the mechanism 
its modern name. Subsequent work showed that a localized bump in surface 
density or pressure would cause the instability \citep{Li00}, and numerical 
simulations \citep{Li01} confirmed its saturation into vortices, 
with most unstable azimuthal wavenumbers $m$=4-5. 

\citet{Varniere&Tagger} raised the possibility that such surface density 
jumps would naturally occur at the boundaries between the magnetized and 
unmagnetized regions of accretion disks. In this boundary, because the 
unmagnetized region constitutes a ``dead zone'' to the MRI \citep{Gammie}, 
there exists a transition in turbulent viscosity. The viscous torque 
at the transition has a component proportional to the negative of the 
viscosity gradient, so material is accelerated outward in the inner 
dead zone boundary (negative viscous gradient) and inward in the outer 
dead zone boundary (positive viscous gradient). This modifies the potential 
vorticity profile at these transitions, triggering the RWI. Because 
this component of the viscous torque is also proportional to the shear rate, 
this process occurs at the shear timescale, not on the much longer viscous timescale. 
 
This two-dimensional scenario, modeling the turbulent transition as jumps 
in alpha-viscosity \citep{Shakura&Sunyaev} was used by \citet{Inaba&Barge}
to argue for accumulation of particles in that region, and by \citet{Lyra08b,Lyra09}
to demonstrate their collapse into planets. The test of this 
scenario in three-dimensional, magneto-hydrodynamical calculations has yet 
to be performed. That is the goal of this paper. 

In particular, we focus on three questions. First, whether the 
RWI will drive angular momentum transport. Previous works \citep{Fleming&Stone,Oishi&MacLow} 
show that although turbulence in the active upper layers 
drives waves that propagate into the dead zone, the amount of stress 
generated is too low to drive significant angular momentum transport, due 
to the low inertia of the upper active layers. In this 
work we investigate the effect of a radial, not vertical, resistive transition. As 
in the midplane the densities are much higher, perhaps the degree of angular momentum 
transport changes significantly. Moreover, strong anti-cyclonic vortices provide 
an exciting possibility. Vorticity generates spiral density waves 
\citep{Heinemann&Papaloizou09a,Heinemann&Papaloizou09b,Heinemann&Papaloizou12}
that would have the right velocity correlation to transport angular momentum outwards. 
It has been shown in local models of unmagnetized disks that by radiation of waves, 
vortices can maintain a relatively high accretion rate, at the level of $\alpha=\ttimes{-3}$ 
\citep{Lesur&Papaloizou,Lyra&Klahr}. \citet{Varniere&Tagger} showed 
proof-of-concept alpha-disk models that the dead zone can be ``revived'' 
by vortices excited in the active/dead boundary, but that still has to be confirmed by MHD 
simulations. 

The second question pertains to the width of the transition. Lyra et al. (2009; see also 
Reg\'aly et al. 2012) show that the maximum width where one expects a pressure bump to 
trigger the RWI is $2H$, where $H$ is the pressure scale height. The outer edge of the 
dead zone undergoes a gradual transition, where the density falls monotonically, until the 
point that it gets so thin that ionizing agents (stellar X-rays or external cosmic rays) 
can penetrate all the way to the midplane. As this 
transition should be very smooth, extending through tens of AU, 
we can exclude that as a possible 
location for RWI excitation. The inner edge, on the other hand, houses
a more exciting possibility, since there the transition in ionization is sharp, set by the collisional 
ionization of potassium, that occurs at $\approx900\,K$ \citep[e.g.][]{Umebayashi,Turner&Drake}.

The other question is whether the 
magnetic field would in any way inhibit the growth of the RWI. \citet{Dzyurkevich} find 
that a density bump develops at the inner edge of the dead zone, but in the timespan of their 
simulation, no appreciable growth of non-axisymmetric modes is observed. The same negative result 
is found in the models of \citet{Kato09,Kato10,Kato12} who model inhomogeneous MRI, 
mimicking the effects of a dead zone. This is curious, since non-axisymmetric modes 
trapped in the pressure bump should develop counter-rotation, as the fluid element ahead of 
the bump is accelerated, and the fluid element behind it is decelerated \citep{Hawley}. 
These negative results could be due to the limited azimuthal extent 
modeled by both works ($\pi/4$ by \citealt{Dzyurkevich}, local box in the works of 
\citeauthor{Kato09}), of stratification in the case of \citet{Dzyurkevich}, or inhibition by 
the magnetic field.

Although vortices have been reported in MRI-active disks 
\citep{Fromang&Nelson05}, the magneto-elliptic instability 
\citep{Mizerski&Bajer,Mizerski&Lyra} was shown by \citet{Lyra&Klahr} to 
be powerful enough to disrupt vortices otherwise stable in non-magnetized 
environments. In this work we present models where the dead zone is modeled by a resistive jump in a magnetized 
disk, and show that the RWI is excited at the dead side of the transition, leading 
to a giant vortex that survives until the end of the simulation. The paper is organized 
as follows. In Sect~2 we present the model equations, and in Sect~3 the initial conditions. 
The results are shown in Sect~4, followed by a discussion and conclusions in Sect~5. 

\begin{figure}
  \begin{center}
    \resizebox{\columnwidth}{!}{\includegraphics{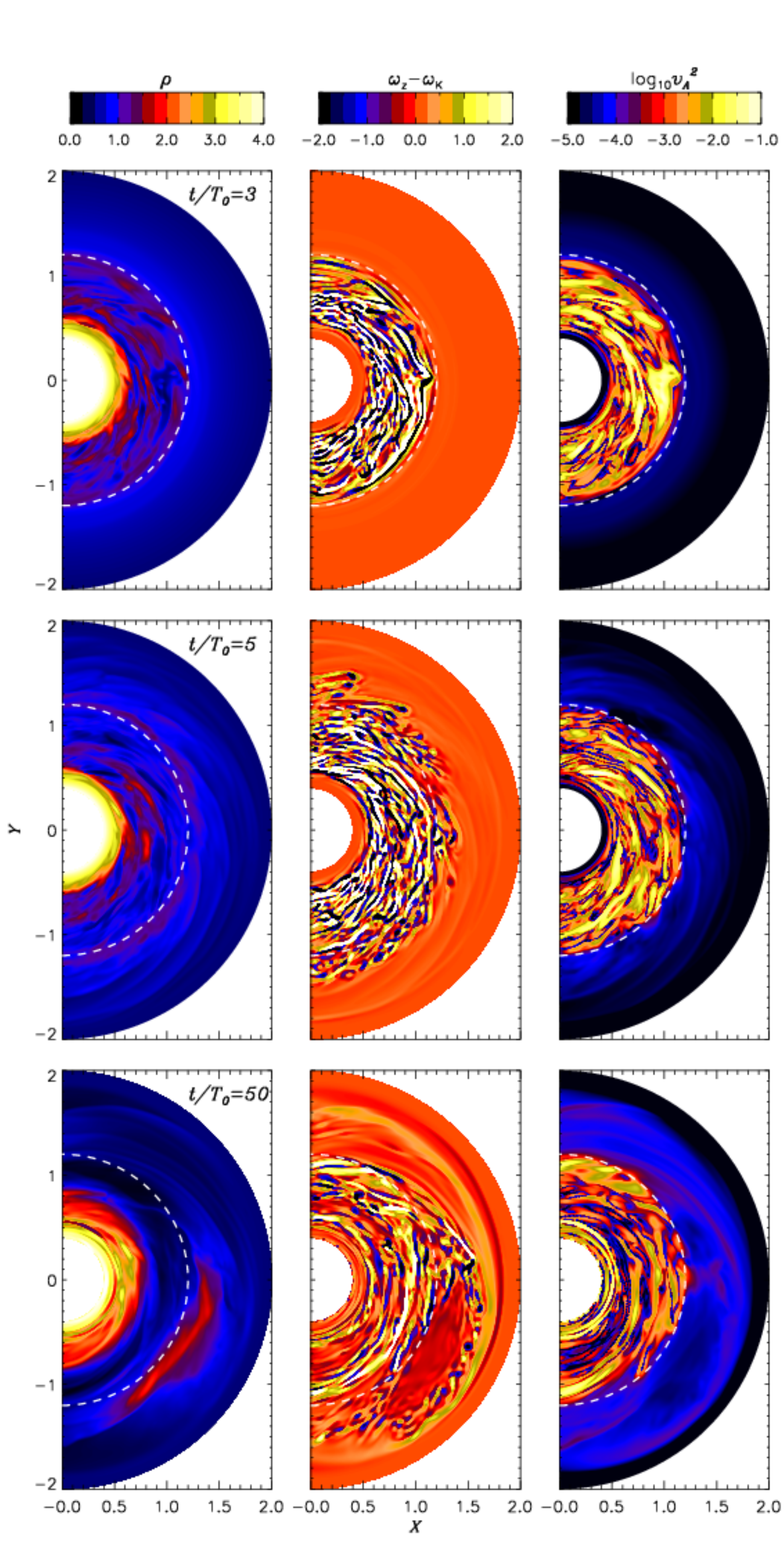}}
  \end{center}
\caption[]{Snapshots of density (left), residual vorticity (middle) and 
Alfv\'en speed (right) at the midplane in selected times at 3, 5, and 50 
reference orbits.}
 \label{fig:fiducial}
\end{figure}

\begin{figure}
  \begin{center}
    \resizebox{\columnwidth}{!}{\includegraphics{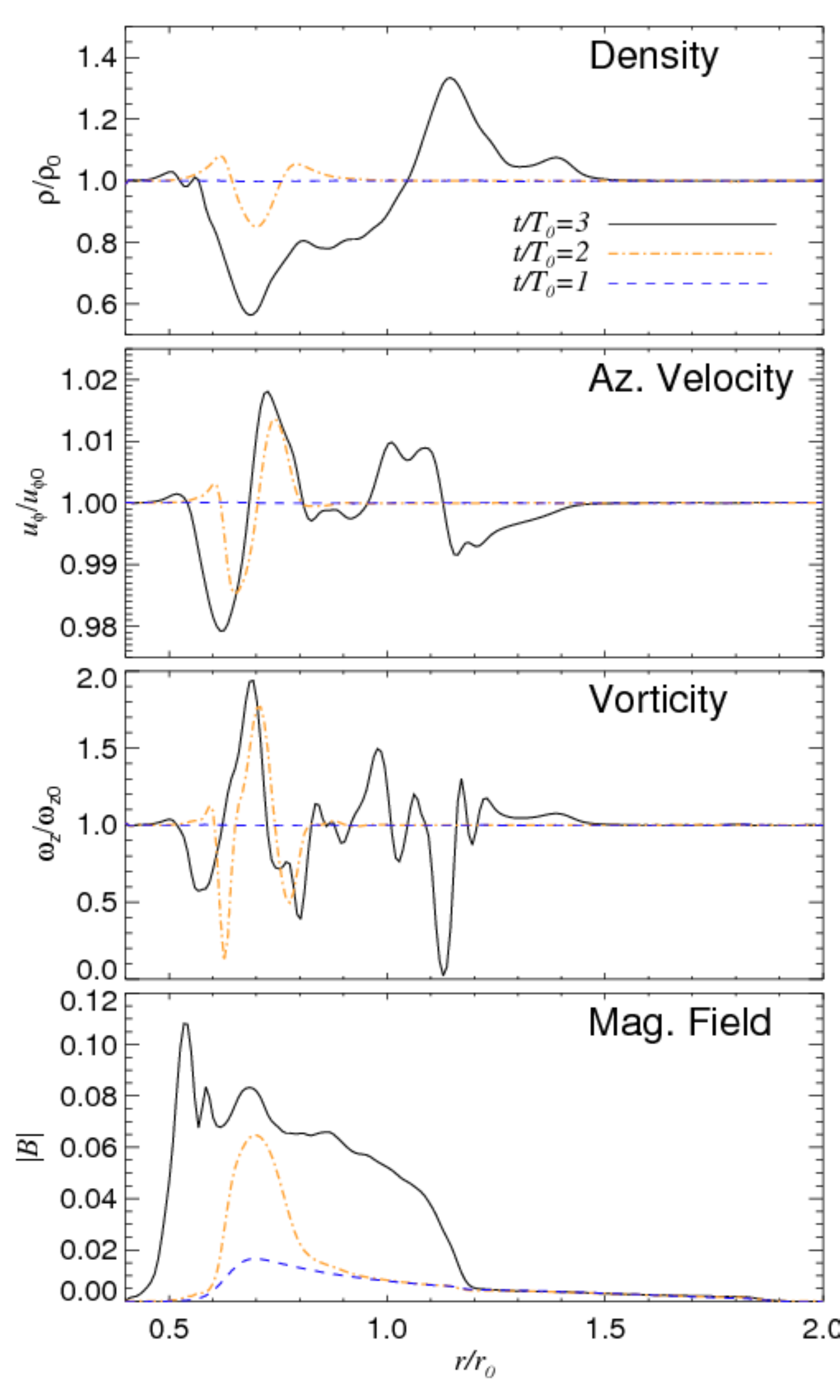}}
  \end{center}
\caption[]{Vertical and azimuthal averages of density, azimuthal 
velocity, vorticity, and magnetic field strength at 1, 2, and 3 orbits, 
i.e., during linear growth of the MRI. The build-up of magnetic 
stress at the resistive transition at $r_\eta=1.2$ gives rise 
to an inversion in azimuthal velocity, with super-Keplerian motion 
inwards and sub-Keplerian motion outwards, generating a dip in 
potential vorticity. The opposite occurs in the artificial peak 
of magnetic pressure at $r=0.7$ coded in the initial condition, leading 
to a peak in potential vorticity. Both locations fulfill the conditions 
for triggering the RWI.}
 \label{fig:rwi-conditions}
\end{figure}

\begin{figure}
  \begin{center}
    \resizebox{\columnwidth}{!}{\includegraphics{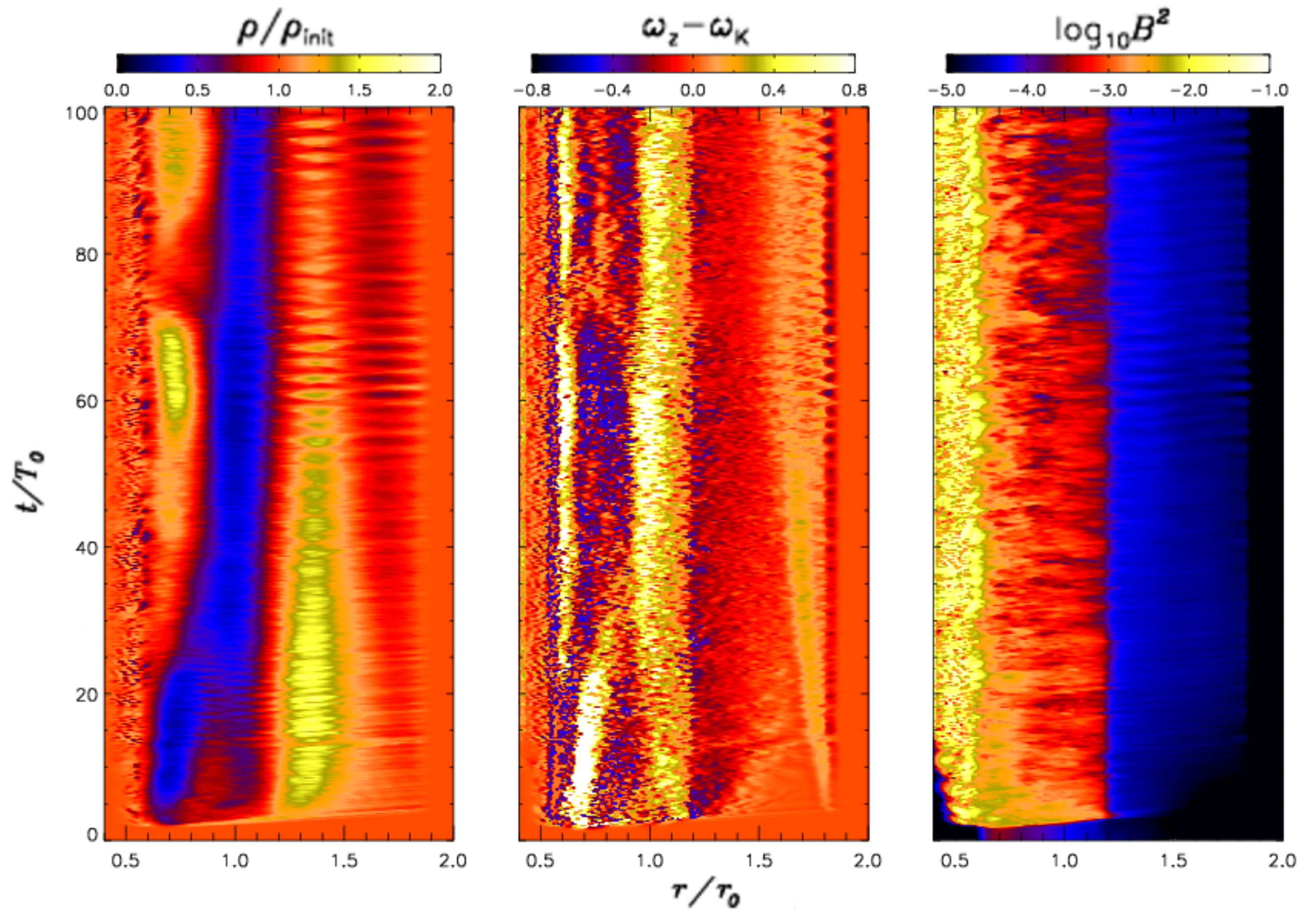}}
  \end{center}
\caption[]{Time evolution of the $\phi$-$z$ averaged ($r$-dependent) 
density (left panel), vertical vorticity (middle panel) and magnetic 
energy (right panel) for the fiducial model. A density enhancement 
is seen at the dead side of the resistive transition at $r=1.2$, 
matching radial locations of lower vorticity. Curiously, 
other vortices are excited in the active zone. One exists at 
$r\approx0.7$ from $t\approx 3 T_0$ to $t\approx 20 T_0$, and another 
next to the inner boundary from $t\approx 40 T_0$ to $t\approx 70T_0$, 
after which it decays and restarts at $t\approx 85T_0$.}
 \label{fig:fiducial-spacetime}
\end{figure}

\section{The model}
\label{sect:model}

\subsection{Dynamical equations}

We perform three-dimensional MHD simulations in the cylindrical 
approximation, i.e., neglecting the disk vertical stratification 
and switching off gravity in that direction. The equations solved are 

\begin{eqnarray}
  \pderiv{\rho}{t}  &=& -\advec\rho -\rho{\Div\v{u}}, \label{eq:continuity}\\
  \pderiv{\v{u}}{t} &=& -\advec\v{u} -\frac{1}{\rho}\del{p} - \del\varPhi + \frac{\v{J}\times\v{B}}{\rho}, \label{eq:navier-stokes}\\
  \pderiv{\v{A}}{t} &=& \v{u}\times\v{B} -\eta\mu_0\v{J} \label{eq:induction}\\
  p&=&\rho c_s^2\label{eq:eos}.
\end{eqnarray}

\noindent where $\rho$ is the density, $\v{u}$ the velocity, $\v{A}$ is the 
magnetic potential, $\v{B}=\curl{\v{A}}$ is the
magnetic field, $\v{J}=\mu_0^{-1}\curl{\v{B}}$ is the current density, and $p$ is the 
pressure. The equation of state is locally isothermal. The gravitational potential $\varPhi=-GM_\star/r$ where 
$G$ is the gravitational potential, $M_\star$ is the stellar mass, and 
$r$ is the cylindrical radius. The resistivity is a radial function of 
position. We use a smooth step function 

\begin{equation}
  \eta(r) = \frac{\eta_0}{2}\left[1+\tanh\left(\frac{r-r_\eta}{\Delta\,r}\right)\right], \label{eq:eta-jump}
\end{equation}

\noindent in order to mimic the effect of a dead zone. The resistivity passes 
from $\eta_0$ to zero over a width $\Delta\,r$ centered at an arbitrarily 
chosen distance $r_\eta$. 

We solve the equations with the {\sc Pencil Code} {\footnote{The code,
including improvements done for the present work, is publicly
available under a GNU open source license and can be downloaded at
http://www.nordita.org/software/pencil-code}} which integrates
the evolution equations with sixth order spatial derivatives, and
a third order Runge-Kutta time integrator. Sixth-order
hyper-dissipation terms are added to \eq{eq:continuity}-\eq{eq:induction},
to provide extra dissipation near the grid scale, explained in
\citet{Lyra08a}. They are needed because the high order scheme of
the Pencil Code has little overall numerical dissipation 
\citep{McNally}.

\subsection{Initial Conditions}

We model a three-dimensional disk on a uniformly spaced mesh in
cylindrical coordinates $(r,\phi,z)$, ranging over $r$=[0.4,2.0]$r_0$
and $z$=[-0.1,0.1]$r_0$, where $r_0$ is a reference radius. 
We run models with azimuthal coverage $L_\phi$=$\pi$ and 
$L_\phi$=$2\pi$. The fiducial model with $L_\phi$=$\pi$ has 
resolution [$N_r$,$N_\phi$,$N_z$]=[192,384,64]. 

The density and sound speed are set as radial power-laws 

\begin{equation}
  \rho = \rho_0 \left(\frac{r}{r_0}\right)^{-q_\rho}; \quad \quad c_s^2 = c_{s0}^2 \left(\frac{r}{r_0}\right)^{-q_{_T}} 
\end{equation}

\noindent with $q_\rho=1.5$ and $q_{_T}=1.0$. The initial angular velocity profile 
is corrected by the thermal pressure gradient

\begin{equation}
  \dot\phi^2 = \varOmega^2 + \frac{1}{r\rho}\frac{\partial{p}}{\partial{r}}
  \label{eq:centrifugal}
\end{equation} \noindent where $\varOmega = \varOmega_0\,(r/r_0)^{-q}$ with $q=1.5$ 
is the Keplerian angular velocity.

The magnetic field is set as a net vertical field, with four MRI wavelengths 
resolved in the vertical range. The constraint $\lambda_{\rm MRI} = 2\pi\va\varOmega^{-1} = L_z/4$ 
translates into a radially varying field 

\begin{equation}
  \va = \frac{L_z\varOmega}{8\pi}\sqrt{\mu_0 \rho} = \vao \left(\frac{r}{r_0}\right)^{-(q+q_\rho/2)},
\end{equation}i.e., a field falling as a 9/4 power-law. We use units such that

\begin{equation}
  GM_\star = r_0 = \rho_0 = \mu_0 = 1 \label{eq:units}
\end{equation}

\noindent and omit these factors hereafter. Thus, $\vao \approx \xtimes{8}{-3}$ 
in code units. The reference sound speed is set at $c_{s0}$=0.1. The dimensionless 
plasma beta parameter $\beta = 2c_s^2/\va^2$ then ranges from 50 
in the inner disk to 1250 in the outer. Noise is added to the velocities 
at the $\ttimes{-4}$ level. In this configuration, the MRI grows and saturates 
quickly in 3 local orbits. We apply the noise and the magnetic field only in the radial 
range $r=$[0.6,1.8] to avoid growth of the instability near the boundaries. We use 
reflective boundaries, with a buffer zone of width 0.2 at each radial border, that 
drives the quantities to the initial condition on a dynamical timescale.

In the presence of resistivity, the excitation of the MRI is controlled 
by the Els\"asser number

\begin{equation}
  \varLambda = \frac{\lambda\va}{\eta},
\end{equation}

\noindent which is the magnetic Reynolds number with velocity equal to the Alfv\'en speed, 
and $\lambda$ the relevant magnetic length scale. We set the reference 
resistivity $\eta_0$ so that the Els\"asser number of the largest 
wavelength present in the box (i.e., $\lambda=L_z$) is unity at $r_0$, 
thus quenching the MRI outward of that radius. This constraint translates into $\eta_0 = L_z \vao = \xtimes{1.6}{-3}$. We 
place the resistivity jump at $r_\eta$=1.2, so that at that location $\varLambda=0.75 < 1$.

\section{Results}
\label{sect:results}

\begin{figure*}
  \begin{center}
    \resizebox{\textwidth}{!}{\includegraphics{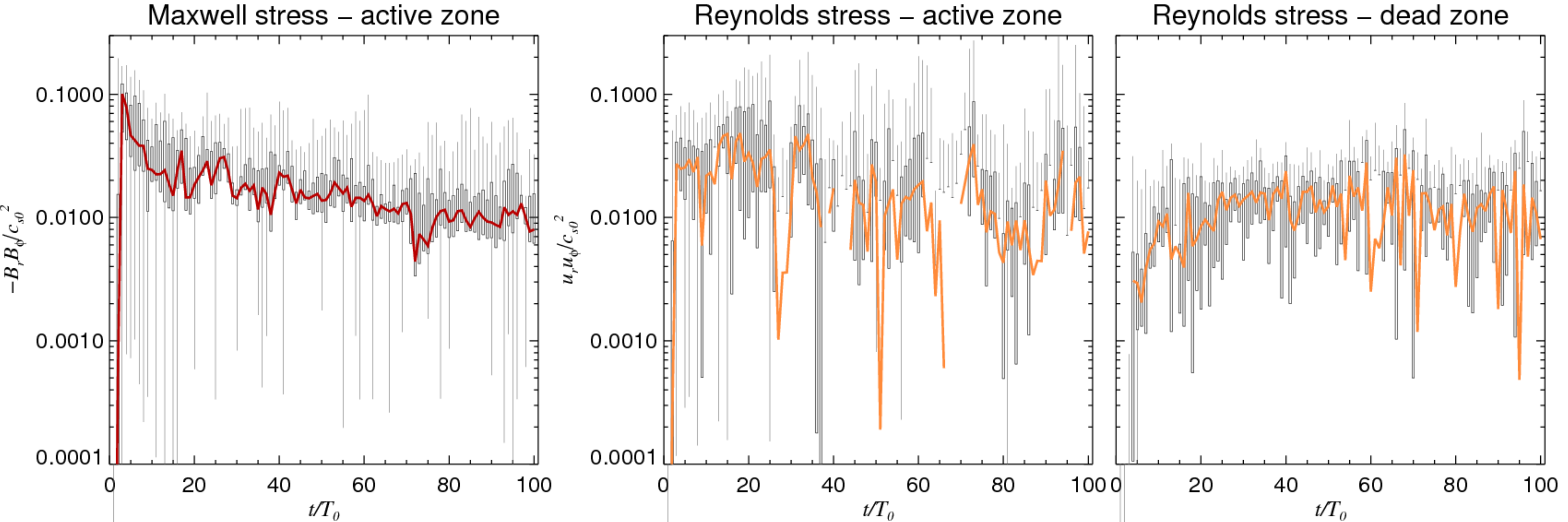}}
  \end{center}
\caption[]{Maxwell and Reynolds stresses as a function of time 
in the active and dead zones. Measurements were taken every 
$T_0=2\pi$, and shown in the plot as box-and-whiskers five point
summaries. The light thin grey lines are the whiskers marking the minimum 
and maximum values, the dark grey thick lines marks the 
lower and upper quartiles that box 50\% of the values. 
A thicker colored line traces the median. High levels of 
Reynolds stress are maintained in the dead zone by the spiral 
density waves excited by the vortex at the active/dead boundary.}
 \label{fig:stresses}
\end{figure*}

\subsection{Fiducial model}

The fiducial model ranges $\pi$ in azimuth, with a resolution 
of [$N_r$,$N_\phi$,$N_z$]=[192,384,64], and a sharp resistivity 
transition at $r_\eta=1.2$ of width $\Delta{r}=\ttimes{-2}$. The MRI 
starts from the inner disk and quickly saturates in 3 local orbits. 
We show in \fig{fig:fiducial} the state of the disk for density (left 
panels), residual vertical vorticity (vertical vorticity minus the vertical 
Keplerian vorticity, middle panels) and Alfv\'en speed (right panels) in 
selected snapshots. The dashed line marks the boundary between 
MRI-active and dead zones. We see that the magnetic energy is well 
confined to the active zone, with only some very minor diffusion into 
the dead zone. The turbulence in the active zone propagates spiral 
density waves into the dead zone as seen in the snapshots at $t/T_0=5$ 
(where $T_0=2\pi$ is the orbital period at $r_0$). 

In \fig{fig:rwi-conditions} we show that the conditions for 
the development of the RWI are fulfilled in this disk model. We plot 
the azimuthal and vertical averages as a function of radius of 
density, azimuthal velocity, vorticity, and magnetic field, for the 
first three orbits. In a barotropic disk, an extremum in potential 
vorticity $\omega_z/\rho$ will launch the RWI \citep{Lovelace}. 
This condition is brought about by the transition in magnetic pressure near 
$r_\eta$.

At later times (lower panels of \fig{fig:fiducial}), a giant vortex is seen on 
the dead side of the transition. The density enhancement (lower left) matches spatially 
a vorticity minimum, confirming its anticyclonic nature. 

We show in \fig{fig:fiducial-spacetime} the time evolution of the 
azimuthal and vertical averages of the same quantities shown in 
\fig{fig:fiducial}. In the density plot we see the vortex as an 
enhancement at the dead size of $r_\eta$. It weakens with time but 
attains a steady state after $t/T_0$=60. The vorticity plot shows 
that the density enhancement is traced by anticyclonic vorticity. 
Conversely the density dips are spatially correlated with regions of 
cyclonic vorticity. Though the resistivity impedes the growth of the 
MRI in the dead zone, it experiences diffusion of some magnetic 
field from the active zone. It thus appears weakly magnetized instead 
of completely demagnetized. 

Interestingly, in the same plots we see that vortices are excited in the 
magnetized regions. A weak vortex is seen in the middle of the active zone 
at very early times, $t/T_0<5$. It eventually decays, after 20 orbits. Other 
intermittent vortices are also seen to be excited, at $r$=0.7, between 40 and 
70 orbits, and again after 85 orbits. These vortices exist in the midst of 
MRI turbulence, as in the models of \citet{Fromang&Nelson05}.
Strictly speaking, \citet{Lyra&Klahr} state that vortices 
that are excited by the baroclinic instability and survive in 
hydrodynamical models are destroyed when 
a magnetic field is abruptly introduced in the simulation. They do not 
exclude excitation of MHD vortices by other ways. Still, these magnetized 
vortices should host magneto-elliptic instabilities \citep{Mizerski&Bajer,Mizerski&Lyra}, 
and given their strength, should not exist; unless vorticity injection by the RWI is 
faster than destruction by the magneto-elliptic instability in the relevant scales. 

The origin of the vortex at $r$=0.7 
lies in the initial condition, because we used a field with a peak in magnetic 
pressure around that radius. This is seen in \fig{fig:rwi-conditions} 
as the peak in magnetic pressure translates into an azimuthal velocity 
inversion and a peak of potential vorticity. This incurs a non-equilibrium configuration, 
that induces sub-Keplerian motion inwards of the peak, and super-Keplerian 
motion outwards. It triggers a localized zonal flow, that in turns excites 
the non-axisymmetric modes of the RWI. 

\subsection{Angular momentum transport}

We display in \fig{fig:stresses} the measured 
Maxwell and Reynolds stresses. 
The Reynolds stress is shown for both the active and dead zone. 
We see that the Maxwell and Reynolds stresses in the active zone 
reach the $10^{-2}$ level, agreeing with previous studies 
\citep{Papaloizou&Nelson,Fromang&Nelson06,Lyra08a,Dzyurkevich,Beckwith,Flock,Sorathia}. 
The novel result of our model is that the Reynolds stress 
in the dead zone is also at the $10^{-2}$ level, 
only slightly lower than in the active zone. 

Without viscous dissipation, the angular momentum is transported 
to the outer boundary, where it is damped, while 
the matter accretes.
As a result of outward angular momentum transport, the vortex 
should migrate inwards \citep{Paardekooper,Lyra&Klahr}
However, this is not observed in the simulations. The vortex keeps 
radiating waves, yet it sits in the pressure maximum that generated 
it, which acts as a migration barrier. The same is seen in the models 
of \citet{Meheut10,Meheut12b}. 

\subsection{Resolution study}

\begin{figure}
  \begin{center}
    \resizebox{\columnwidth}{!}{\includegraphics{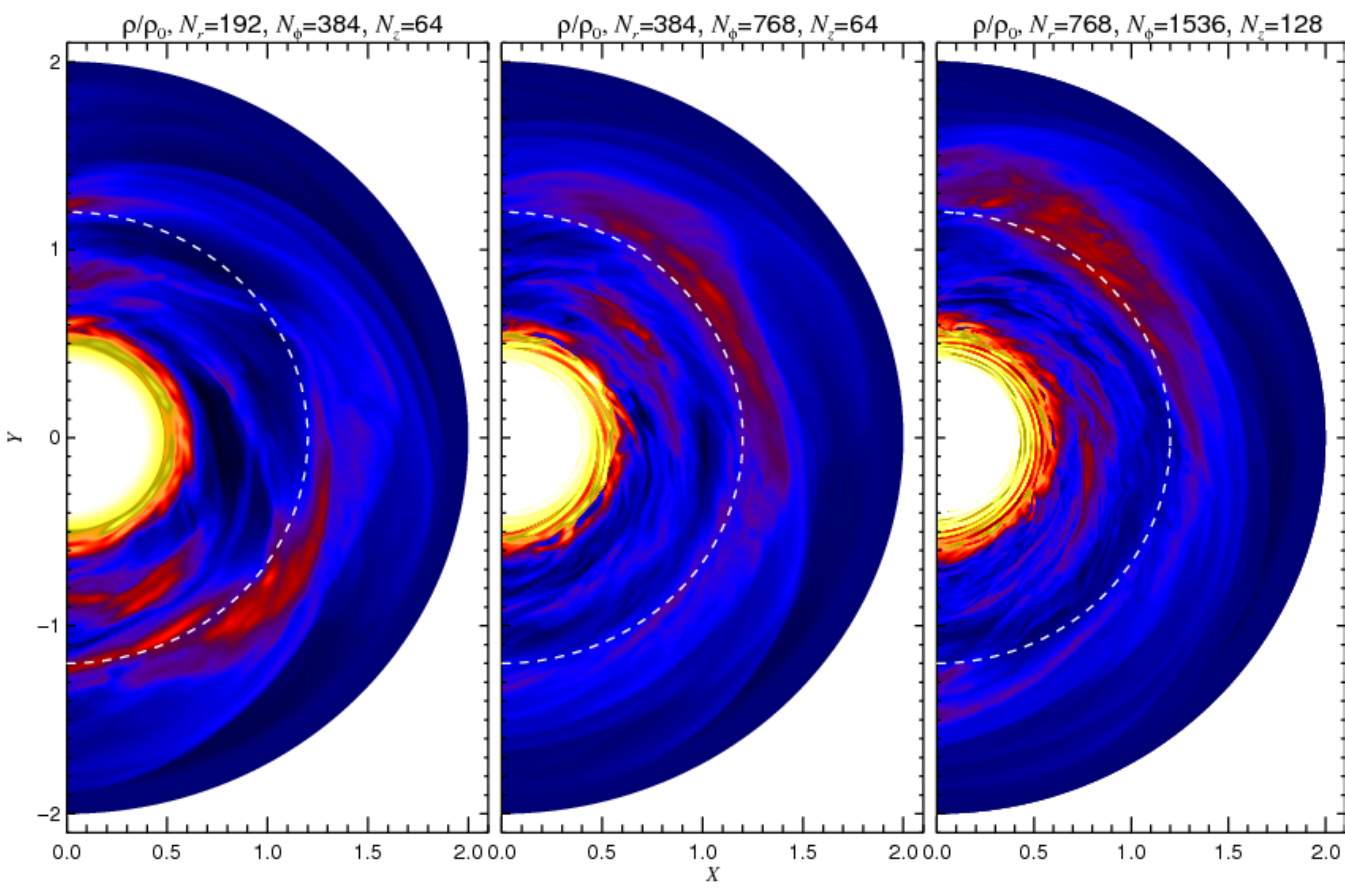}}
  \end{center}
\caption[]{Models of resolution 192$\times$384, 384$\times$768 
and 768$\times$1536 in the midplane. The vertical resolution 
is 64 in the first two plots, and 128 in the last one. 
The vortex in the magnetized 
zone, that should be unstable, shrinks in size with increasing 
resolution, but is not quenched even in the highest resolution 
model (see text). The outer vortex exists in the dead zone and 
is not subject to the same destabilizing mechanism. The same 
colorbar is used as in \fig{fig:fiducial}.}
 \label{fig:resolution}
\end{figure}

\begin{figure*}
  \begin{center}
    \resizebox{.75\columnwidth}{!}{\includegraphics{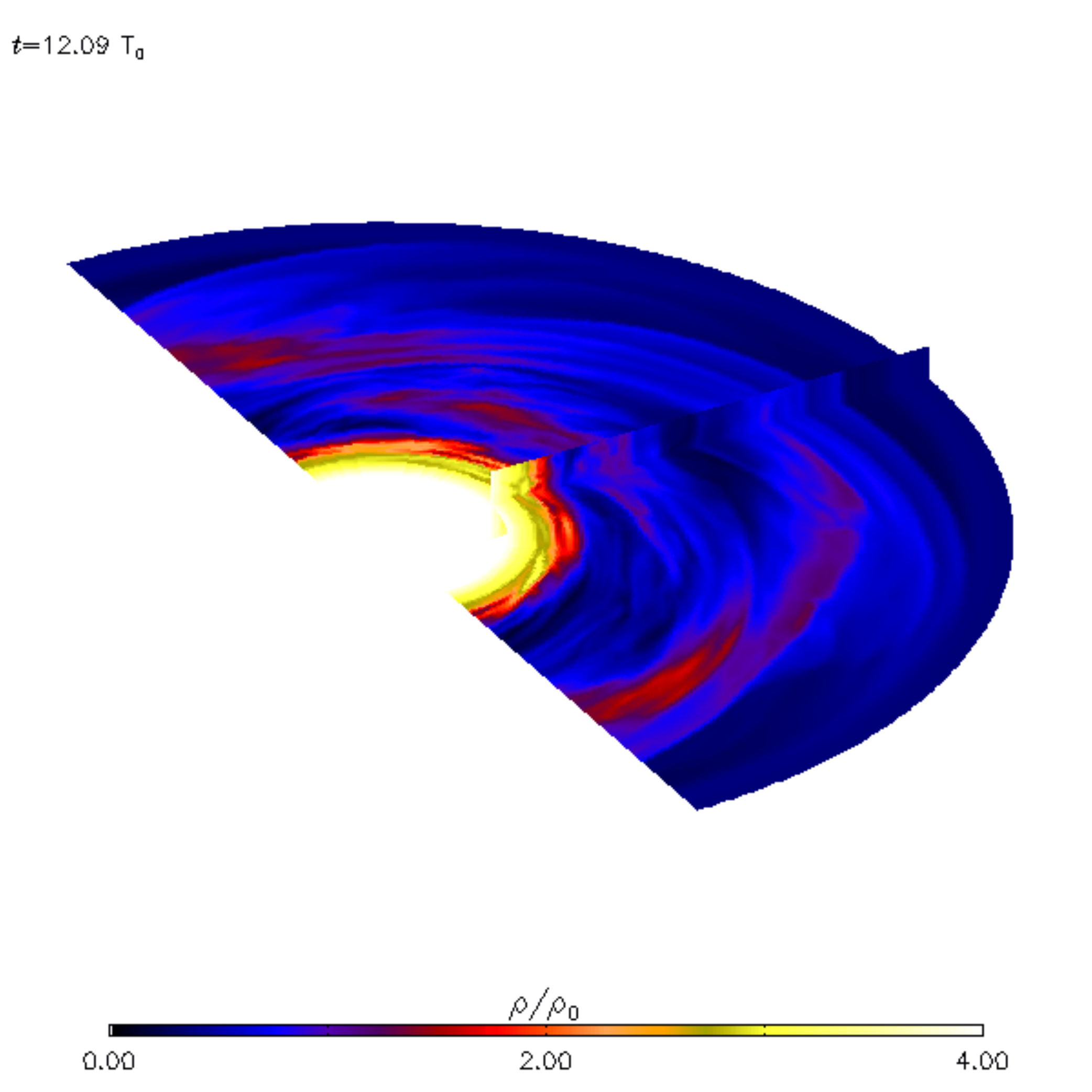}}
    \resizebox{.75\columnwidth}{!}{\includegraphics{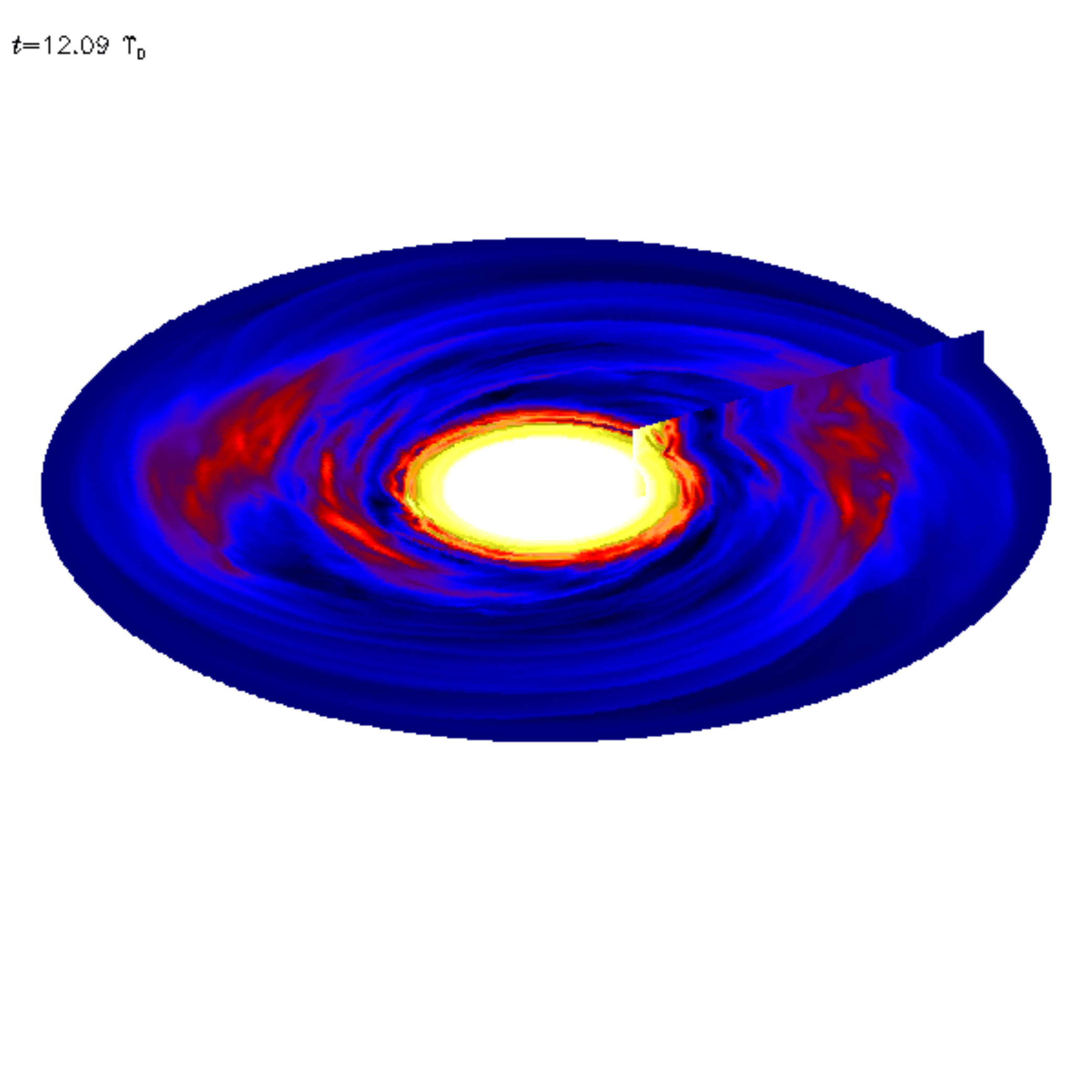}}
    \resizebox{.75\columnwidth}{!}{\includegraphics{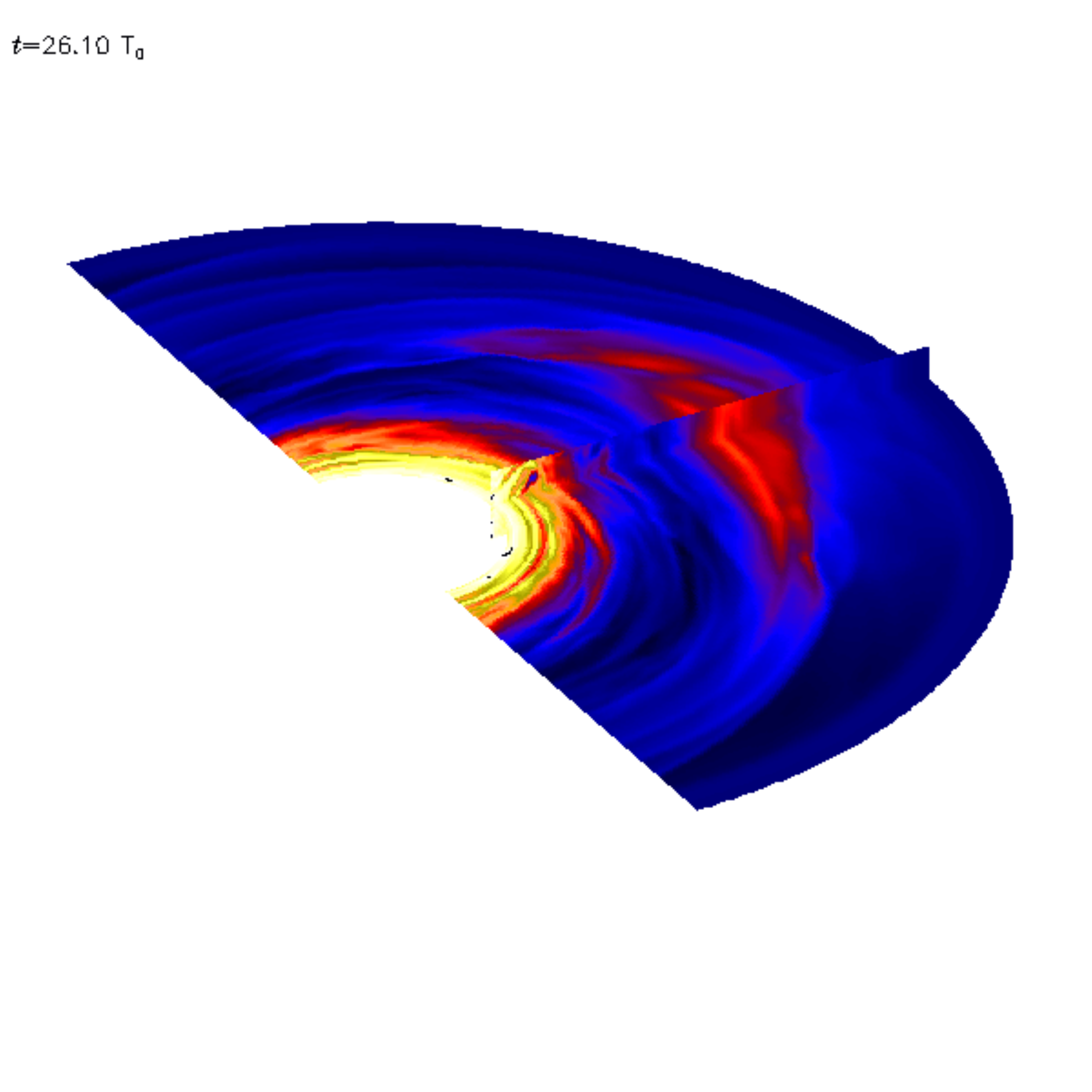}}
    \resizebox{.75\columnwidth}{!}{\includegraphics{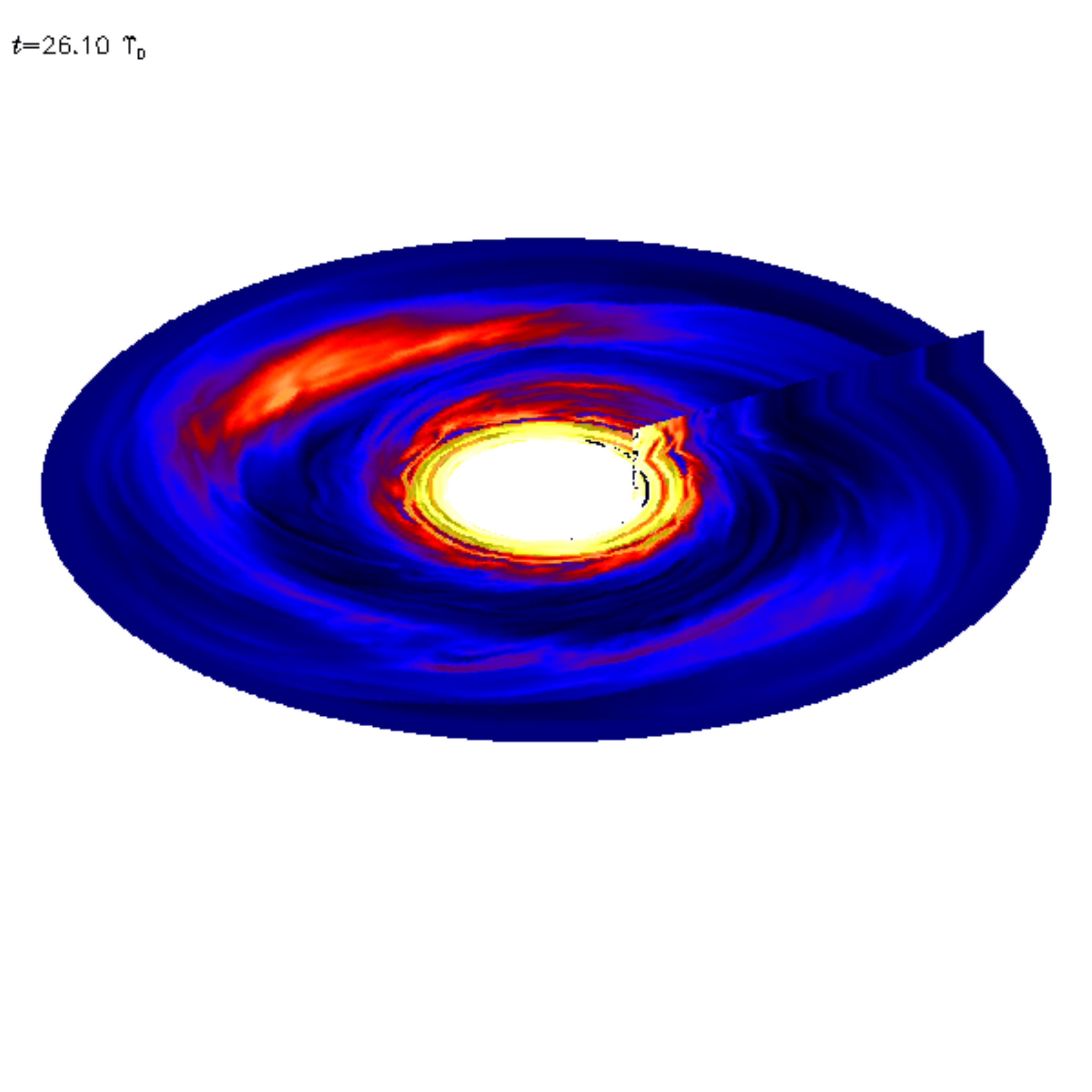}}
  \end{center}
\caption[]{Since the RWI is non-axisymmetric, we assess 
the influence of azimuthal domain size $L_\phi$ on its 
evolution by comparing models of $L_\phi=\pi$ (left panels) 
and $L_\phi$=$2\pi$ (right panels). Upper panels 
show the models at 12 orbits, lower panels at 26 orbits. Despite
differences at early times (at 12 orbits the vortex cascade is 
still at $m=2$ in the $2\pi$ models), once the inverse cascade 
is complete and a single vortex survives, the models look 
remarkably similar.}
 \label{fig:2pi}
\end{figure*}

\begin{figure*}
  \begin{center}
    \resizebox{\textwidth}{!}{\includegraphics{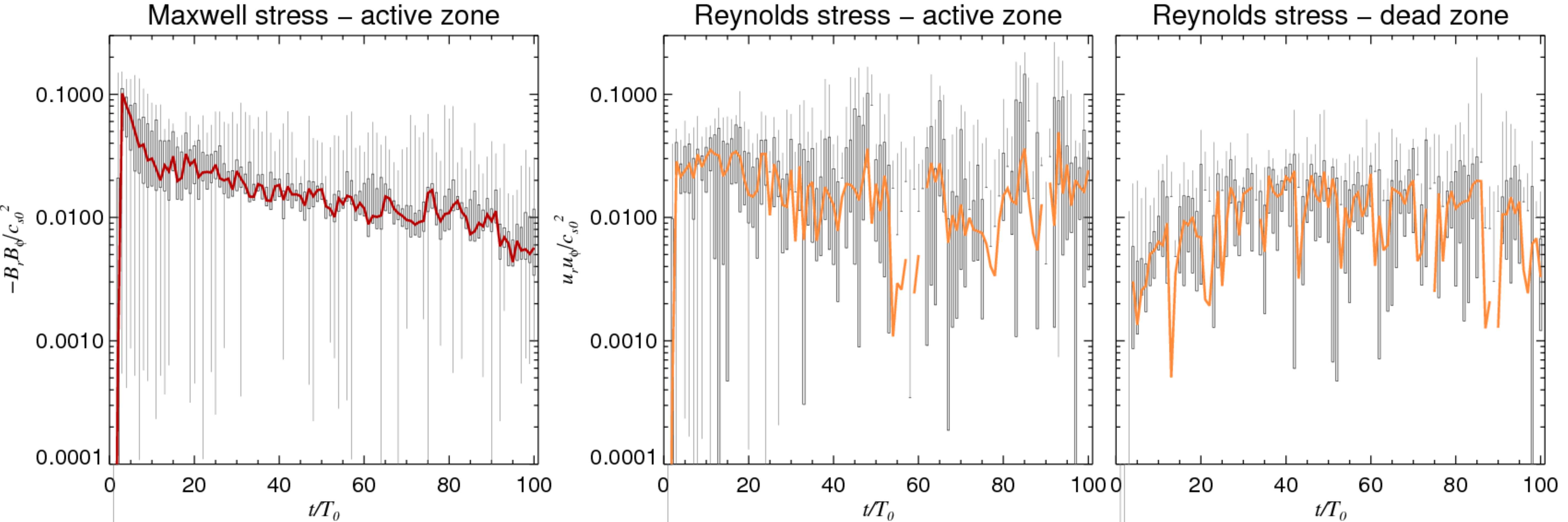}}
  \end{center}
\caption[]{Same as \fig{fig:stresses} but for the full 2$\pi$ model. The level
of angular momentum transport is very similar to the model spanning only $\pi$ 
in azimuth. We conclude that the latter model leaves enough azimuthal room 
to avoid vortex self-interaction across the periodic boundary.}
 \label{fig:stresses-2pi}
\end{figure*}

According to the results of \citet{Lyra&Klahr}, the magneto-elliptic 
instability should not allow the formation of the vortices seen in the 
MRI-active region. This suggests that the models presented 
are under-resolving the MEI-modes. To examine this possibility, we run 
models with twice ($N_r=384$, $N_\phi=768$, $N_z=64$) and four times 
($N_r=768$, $N_\phi=1536$, $N_z=128$) the midplane resolution of the 
fiducial model. The flow pattern after 12 orbits is shown in 
\fig{fig:resolution}. We see that the inner vortex in the magnetized 
region has shrunk in size while the outer one in the dead zone has 
not. This is evidence that the MEI is being excited, and the vortex core 
shrinks to a size where it is under-resolved. The size of the scale height 
in all models is $H=hr=0.1\,r$, so $H$=0.08 for the location of the 
inner vortex. The low resolution model resolves that scale with 
$H/\Delta\,r$=9 points, the middle resolution one with 19 points 
and the highest resolution model with 38 points. 

We see that even in the highest resolution model, the density enhancement 
persists. We conclude that this vortex is either magnetically stable 
(which is problematic in view of the claims of \citealt{Lyra&Klahr})
or the resolution requirement to capture the unstable modes has not been met. 
As we could not push the resolution even further (the highest resolution 
model already consumed 2 million computer hours), 
the question of the stability of the inner vortex will be addressed 
in a future work, in local box models, to satisfy the resolution 
requirement. 

\subsection{$2\pi$ model}

As shown by \citet{Flock} the main features of the MRI can be captured 
with a domain spanning $\pi$ in azimuth as well as they can in a domain 
spanning $2\pi$. That statement, however, though applying for the 
axisymmetric MRI, does not automatically apply to the non-axisymmetric 
RWI, for which the most unstable wavelengths are $m=4$ and $m=5$ \citep{Li00}. 
Viscous mergings then transfer the power towards lower wavenumbers, 
in a cascade leading to $m=1$, i.e., a single vortex{\footnote{
The $m=1$ mode can cascade further, back to $m=0$, thus turning the 
RWI into a cycle \citep[see][]{Regaly}. However, this development occurs 
only over long evolution times and will not be dealt 
with in this work.}}.

We check if there are significant differences pertaining to azimuthal 
domain size by performing a simulation spanning 2$\pi$ radians 
and comparing the result 
with the fiducial model. The model has the same numerical resolution, 
using twice the grid points in $\phi$, [$N_r$,$N_\phi$,$N_z$]=[192,768,64].
The comparison is shown in \fig{fig:2pi}. It 
is seen that at 12 orbits the $2\pi$ model shows a conspicuous $m$=2 mode, 
whereas the $\pi$ model contains a single vortex at the same time. The 
two vortices in the $2\pi$ model later merge into a $m=1$ mode, and as 
a result the flows in both cases are very similar at $t$=26 orbits.

We also measure the angular momentum transport in the $2\pi$ model, shown 
in \fig{fig:stresses-2pi}. The stresses are very similar to the ones 
shown in \fig{fig:stresses}, for the $\pi$ model. This means a box size 
of $L_\phi$=$\pi$ already provides enough azimuthal spacing so that the 
vortex does not interact with itself in a way that significantly affects 
the stresses (the same is not true for local boxes). 

We conclude that studies concerning the linear growth of the instability 
require full $2\pi$ simulations, yet $\pi$ suffices if the interest is in the 
saturated, $m=1$, state. 

\section{Conclusions}
\label{sect:conclusions}

We performed the first simulation of a RWI-unstable boundary between 
dead and active zones in a protoplanetary disk with 3D 
resistive-magnetohydrodynamical models. We use a sharp resistive 
transition, that should be characteristic of the inner dead zone boundary. 

We find that Rossby vortices are excited on the 
dead side of the transition. They first appear in an $m=4$ mode, 
as expected from the RWI, then quickly merge viscously 
into a $m=2$ mode, and finally reach $m=1$, becoming 
a single giant vortex 
immediately outside of the transition. This lends credence to 
previous works that relied on alpha disk models with angular velocity 
transitions to trigger the RWI \citep{Inaba&Barge,Lyra08b,Lyra09,Meheut10,Meheut12b}.

We also assess the mass accretion rate in the dead zone, which is a 
result both of waves propagating from the active zone and of waves 
radiated by the vortex. We find normalized Reynolds stresses of 
similar magnitude in both the active and in the dead zones, at the 
$\ttimes{-2}$ level. Vortex-free transitions develop dead zone 
stresses two orders of magnitude quieter 
than the active zones, so this high accretion rate is a manifestation 
of the ability of the vortex to maintain high accretion rates. 
The same level is found for angular momentum transport in 
baroclinically unstable local 
disk models \citep{Lesur&Papaloizou,Lyra&Klahr}. Because the 
vortex is born in a high pressure ridge, it does not migrate 
further in; the ridge acts as a migration barrier. 

A puzzling aspect of our model is the development of a weak 
vortex in the middle of the active zone. The vortex was driven
by the RWI at the peak in magnetic pressure due to the artificial 
initial condition. Yet according to a recent study by 
\citet{Lyra&Klahr}, rotating magnetized vortices should be host to 
a set of unstable centrifugal hydromagnetic modes (magneto-elliptic 
instability) that should ultimately break closed elliptic streamlines.
If that is the case, then our models are either under-resolving these modes, 
or vorticity injection by the RWI is faster than destruction by the magneto-elliptic 
instability. We pushed the resolution of the global model to the limit allowed 
by current computational power. The vortex shrank with increasing 
resolution, but we could not quench it completely. A conclusive 
test will require local models, which we plan to undertake in future 
work. A study of convergence with dead zone size, and an assessment of 
how wide the resistive jump can be while still allowing for Rossby vortex 
excitation are also lacking. 

The models presented leave room for exciting developments. 
First, we used 
a static resistivity profile. Though this is shown by \citet{Latter&Balbus}
to produce acceptable results, we plan to include dynamic resistivity in 
future models, following the reduced recombination network of 
\citet{Ilgner&Nelson06a,Ilgner&Nelson06b,Ilgner&Nelson06c}, 
used by \citet{Turner&Sano,Turner&Drake,Gressel11,Hirose&Turner} and \citet{Gressel12}.
Effects of Hall MHD and ambipolar diffusion \citep{Wardle,Salmeron&Wardle,Bai&Stone,Bai} 
may also impact significantly on the nature of the 
active/dead radial transition. Future work will also include 
stratification, shown through numerical \citep{Meheut10,Meheut12b} and 
analytical work \citep{Meheut12a,Lin}
to matter significantly for the internal dynamics 
of 3D vortices. Finally, interacting particles should be included 
to assess the ability of three-dimensional Rossby vortices 
in dead zones to assist on planet formation. 

\acknowledgments 

The writing of this paper was started at the American 
Museum of Natural History with financial support by the National Science 
Foundation under grant no. AST10-09802, and completed at the Jet Propulsion 
Laboratory, California Institute of Technology, under a contract with the 
National Aeronautics and Space Administration. This work was performed 
under allocation TG-MCA99S024 from the Extreme Science and Engineering 
Discovery Environment (XSEDE), which is supported by National Science 
Foundation grant number OCI-105357. The computations were performed on 
the Kraken system at the National Institute for Computational Sciences.
We thank the anonymous referee for useful suggestions.

\end{document}